\documentclass[preprint,12pt]{elsarticle}
\usepackage{amssymb}

\newcommand{\omgpe}{\omega_{\mathrm{p}e}}
\newcommand{\omgps}{\omega_{\mathrm{p}s}}
\newcommand{\omgce}{\omega_{\mathrm{c}e}}
\newcommand{\kv}{\mathbf{k}}
\newcommand{\vv}{\mathbf{v}}
\newcommand{\xv}{\mathbf{x}}
\newcommand{\kDe}{k_{\mathrm{D}e}}
\newcommand{\kDs}{k_{\mathrm{D}s}}
\newcommand{\vthe}{v_{\mathrm{th},e}}
\newcommand{\vths}{v_{\mathrm{th},s}}
\newcommand{\sgn}{\mathrm{sgn}}
\newcommand{\kmax}{{k_\mathrm{max}}}
\newcommand{\kmin}{{k_\mathrm{min}}}

\newcommand{\lmde}{\lambda_e}
\newcommand{\Dx}{{\Delta x}}

\begin{document}

\date{August 12, 2013}

\begin{frontmatter}



\title{Energy Loss of High-Energy Particles in Particle-in-Cell Simulation}


\author{Tsunehiko N. Kato\corref{cor1}}
\cortext[cor1]{Corresponding author.}
\ead{kato@hiroshima-u.ac.jp}

\address{Department of Physical Science, Hiroshima University, 1-3-1 Kagamiyama, Higashi-Hiroshima, Hiroshima 739-8526, Japan}

\begin{abstract}
When a charged particle moves through a plasma
at a speed much higher than the thermal velocity of the plasma,
it is subjected to the force of the electrostatic field induced in the plasma by itself
and loses its energy.
This process is well-known as the stopping power of a plasma.
In this paper we show that
the same process works in particle-in-cell (PIC) simulations as well
and the energy loss rate of fast particles due to this process is mainly determined by the number of plasma electrons
contained in the electron skin depth volume.
However,
since there are generally very few particles in that volume in PIC simulations
compared with real plasmas,
the energy loss effect can be exaggerated significantly and can affect the results.
Therefore,
especially for the simulations that investigate the particle acceleration processes,
the number of particles used in the simulations
should be chosen large enough to avoid this artificial energy loss.
\end{abstract}


\end{frontmatter}

\section{Introduction}
Recently,
with the growth of the computational power,
large-scale PIC simulations have been performed
for many purposes.
These simulations, for example,
provide a direct method to investigate
the particle acceleration processes in collisionless shocks
\cite{Riquelme11, Matsumoto12}.
The acceleration processes
found in such simulations (mainly for electrons) work in relatively short timescales.
On the other hand,
the acceleration processes that work more slowly and need longer timescales,
for instance, the first-order Fermi acceleration in collisionless shocks,
may be affected by the energy loss effect of the high energy particles.
In fact,
as shown in this paper,
the high energy particles can be suffered from
significant energy loss if the number of particles
used in simulation is too small.
In such cases,
the acceleration process can become inefficient or cease completely.

The cause of the energy loss of high energy or fast particles in PIC simulations
is physical, not numerical.
It is essentially the same as
that of the stopping power in a plasma \cite{Gould72}.
However,
since the effects of binary collisions,
ionization of atoms, charge exchange, electron spin,
and other quantum effects, etc.,
are usually not incorporated in the model of PIC simulations,
the energy loss of fast particles
is mainly brought
by the classical plasma response
through the longitudinal electrostatic fields,
or the polarization drag force,
which is described by the dielectric response function of the plasma \cite{Ichimaru}.

The energy loss rate in PIC simulations can also depend on
the dimensionality of simulation.
In the following,
we deal with those in one and two dimensions
as well as in three dimensions.
By one dimension,
we mean here that the configuration space is one-dimensional
while the velocity space is three-dimensional;
this situation is often called as `1D3V'.
Similarly, by two dimensions we mean `2D3V'.
While these plasmas are different from
the \textit{true} one- or two-dimensional (i.e., 1D1V or 2D2V) plasmas,
e.g. \cite{Platzman76},
they are widely used in PIC simulations because
they can be regarded as three dimensional models
with some restrictions or translational symmetries in relevant directions
and they may include more physics of real three-dimensional plasmas
than the true one- or two-dimensional models.

In this paper,
we first derive theoretical expressions for the energy loss rate of fast particles
in PIC simulations
in one, two, and three dimensions in Section 2.
Then,
in Section 3,
several comparisons are made between the theoretical results and numerical experiments.
Finally,
concluding remarks are presented in Section 4.

\section{Energy loss rate}
In this section, we first briefly introduce the dielectric response function
which describes the energy loss rate of fast particles in a plasma
due to the electrostatic response of the plasma based on \cite{Ichimaru}.
Then, we derive the energy loss rates first for three dimensions
and then for one and two dimensions.

Consider a fast particle with mass $m$ and charge $q_0$ moving through a uniform plasma
at velocity $\vv_0$.
Its charge density is given by
\begin{equation}
	\rho(\xv,t) = q_0 \delta(\xv - \vv_0t).
	\label{eq:charge_density_3d}
\end{equation}
Since the electrostatic potential induced by the fast particle in the plasma is not symmetric,
the fast particle is subjected to a force due to the induced field
and decelerated.
The energy loss rate is given by \cite{Ichimaru}
\begin{equation}
	\frac{dE}{dt} = 4\pi q_0^2 \sum_\kv \frac{\kv\cdot\vv_0}{k^2} \mathrm{Im}\left[\epsilon(\kv, \kv\cdot\vv_0)\right]^{-1},
	\label{eq:loss_rate}
\end{equation}
where $k=|\kv|$ and $\epsilon(\kv, \omega)$ is the dielectric response function of the plasma.
The dielectric response function is independent of the spatial dimensionality for homogeneous plasmas,
and hence it is common to 1D3V, 2D3V and three-dimensional simulations.
The summation for $\kv$ in the equation, however, depends on the dimensionality of simulation
and it should be regarded as
\begin{equation}
	\sum_\kv = \frac{1}{(2\pi)^d} \int d^dk,
\end{equation}
where $d$ is the dimensionality considered ($d = 1, 2, 3$).

For a Maxwellian electron plasma with charge neutralizing background,
the dielectric response function is given by
\begin{equation}
	\epsilon(\kv, \omega) = 1 + \frac{\kDe^2}{k^2} W\left(\frac{\omega}{k \vthe}\right),
	\label{eq:DRF}
\end{equation}
where $\vthe = (T_e/m_e)^{1/2}$ is the thermal velocity of the electrons in the plasma
and $\kDe = \omgpe / \vthe$ is the electron Debye wavenumber.
Here,
$T_e$ is the electron temperature,
$\omgpe = (4\pi n e^2/m_e)^{1/2}$ is the electron plasma frequency,
$m_e$ is the electron mass,
$-e$ is the electron charge,
and $n$ is the mean number density of the plasma electrons.
The W function, $W(Z)$, is given in \cite{Ichimaru}
and can be expressed in terms of
the plasma dispersion function $\tilde{Z}(\zeta)$ as
\begin{equation}
	W(Z) = 1 + \frac{Z}{\sqrt{2}} \tilde{Z}\left(\frac{Z}{\sqrt{2}}\right).
\end{equation}
When the argument $Z$ is real, the imaginary part of the W function is given by
\begin{equation}
	\mathrm{Im}[W(Z)] = \left(\frac{\pi}{2}\right)^{1/2} Z \exp\left(-\frac{Z^2}{2}\right).
	\label{eq:W_imag}
\end{equation}
When $Z$ is real and $|Z|= |v_0\mu/\vthe| \gg 1$,
where $v_0 = |\vv_0|$ and $\mu = \kv \cdot \vv_0 / kv_0$,
the real part can be approximated as
\begin{equation}
	\mathrm{Re}[W(Z)] \sim -\frac{1}{Z^2}
	\label{eq:W_approx}
\end{equation}
and with this approximation we have
\begin{equation}
	\mathrm{Im}[\epsilon(\kv, \kv\cdot\vv_0)]^{-1} \sim - \frac{k^2 \varepsilon}{[k^2-(\omgpe/v_0\mu)^2]^2 + \varepsilon^2},
	\label{eq:inv_epsilon_approx}
\end{equation}
where $\varepsilon \equiv (\pi/2)^{1/2}\kDe^2Z\exp(-Z^2/2)$.
Note that for sufficiently small $\varepsilon$
this function (\ref{eq:inv_epsilon_approx}) takes
effectively zero anywhere except around $k = \omgpe/v_0\mu$
and
using the relation for the delta function
\begin{equation}
	\lim_{\varepsilon \to \pm 0} \frac{\varepsilon}{x^2 + \varepsilon^2} = \pm\pi\delta(x)
\end{equation}
it is approximated further as
\begin{equation}
	\mathrm{Im}[\epsilon(\kv, \kv\cdot\vv_0)]^{-1}
	\sim 
	-\frac{\pi \omgpe}{2 k v_0} \sgn(\mu) \left[  \delta(\mu - \omgpe/kv_0) + \delta(\mu+\omgpe/kv_0)\right].
	\label{eq:inv_epsilon_approx2}
\end{equation}
This can be used as a good approximation below
because we consider only sufficiently fast particles ($v_0 \gg \vthe$) in this paper.

\subsection{Three dimensions}
In three dimensions,
the equation (\ref{eq:loss_rate})  becomes
\begin{equation}
	\frac{dE}{dt} = \frac{q_0^2 v_0}{\pi} \int_\kmin^\kmax dk \int_{-1}^1 d\mu \ k\mu \mathrm{Im}\left[ \epsilon(\kv,kv_0\mu)\right]^{-1}.
	\label{eq:loss_rate_3d}
\end{equation}
For PIC simulations, the lower and upper limits of the $k$ integration may be given by
\begin{equation}
	\kmin = \frac{2\pi}{L},
	\qquad \textrm{and} \qquad
	\kmax = \frac{\pi}{\Dx},
\end{equation}
where $L$ and $\Dx$ are the system length and the grid size of the simulation,
respectively.
Using the approximation (\ref{eq:inv_epsilon_approx2})
and assuming $L > 2\pi \lmde$, which is usually satisfied in PIC simulations,
we finally obtain
\begin{equation}
	\frac{dE}{dt} = -\frac{q_0^2 \omgpe^2}{v_0} \ln(\kmax v_0/\omgpe),
	\label{eq:loss_rate_3d_c}
\end{equation}
which is equivalent to the expression of the well-known stopping power
in a plasma.
Introducing the number of electrons contained in the volume of the cubic electron skin depth,
$N_e^{(3)} = n \lmde^3$,
where $\lmde \equiv c/\omgpe$ is the electron skin depth,
the equation (\ref{eq:loss_rate_3d_c}) can be rewritten as
\begin{equation}
	\frac{dE}{dt} = -\frac{(q_0/e)^2}{4\pi N_e^{(3)}} \frac{c}{v_0} \ln(\kmax v_0/\omgpe) \ \omgpe m_e c^2.
	\label{eq:loss_rate_3d_d}
\end{equation}
%

\subsection{One dimension (1D3V)}
The one-dimensional (1D3V) cases can be regarded as three-dimensional ones
but with a restriction or a symmetry;
letting the variable direction be the $x$ direction,
the translational symmetry in $y$ and $z$ directions is required.
Therefore,
even if we want to deal with only one fast particle,
we inevitably must consider
identical fast particles (with the same $x$ coordinate)
uniformly distributed on the $y$-$z$ plane like a charged `sheet'.
Thus,
instead of (\ref{eq:charge_density_3d}),
the charge density of the fast particles is given as that made by the sheet:
\begin{equation}
	\rho(\xv,t) = q_0 \sigma \delta(x-v_{0x}t),
\end{equation}
where $\sigma$ is the two-dimensional number density of the fast particles
embedded on the sheet.
Taking into account of this condition,
the energy loss rate \textit{per particle} (\ref{eq:loss_rate}) for 1D3V cases becomes
\begin{equation}
	\frac{dE}{dt} = 2 q_0^2 \sigma v_{0x} \int \frac{1}{k} \mathrm{Im} \left[\epsilon(k,kv_{0x})\right]^{-1} dk
\end{equation}
with
\begin{equation}
	\epsilon(k,kv_{0x}) = 1 + \frac{\kDe^2}{k^2} W\left(\sgn(k) \frac{v_{0x}}{\vthe}\right).
\end{equation}
The ranges of the integration are given by $[-\kmax, -\kmin]$ and $[\kmin, \kmax]$.
This can be integrated without any approximations to obtain
\begin{equation}
	\frac{dE}{dt} = -2q_0^2 \sigma |v_{0x}| \left[ \arctan \left( \frac{(\kmax/\kDe)^2 + w_r}{|w_i|} \right) - \arctan\left(\frac{(\kmin/\kDe)^2 + w_r}{|w_i|} \right) \right],
	\label{eq:loss_rate_1d_0}
\end{equation}
where
\begin{equation}
	w_r \equiv \mathrm{Re}\left[ W\left(\frac{|v_{0x}|}{\vthe}\right) \right]
	\quad \textrm{and} \quad
	w_i \equiv \mathrm{Im}\left[ W\left(\frac{|v_{0x}|}{\vthe}\right) \right].
	\label{eq:W_re_im}
\end{equation}
Since the argument of the W function, $Z = |v_{0x}|/\vthe$, is real,
$w_i$ is given by the equation (\ref{eq:W_imag}).

For fast particles with $|v_{0x}| \gg \vthe$ (i.e., $Z \gg 1$),
from equations (\ref{eq:W_imag}) and (\ref{eq:W_approx}),
we see that $|w_i| \ll |w_r| \ll 1$.
In addition,
in PIC simulations the conditions
\begin{equation}
	\frac{\kmin}{\kDe} = 2\pi\frac{\lmde}{L} \frac{\vthe}{c} < \frac{\vthe}{|v_{0x}|} = |w_r|^{1/2}
	\quad \textrm{and} \quad
	\frac{\kmax}{\kDe} = \pi \frac{\lmde}{\Dx} \frac{\vthe}{c} > \frac{\vthe}{|v_{0x}|} = |w_r|^{1/2}
\end{equation}
are usually satisfied for fast particles.
With these conditions,
we finally obtain
\begin{equation}
	\frac{dE}{dt} = -2 \pi q_0^2 \sigma |v_{0x}|.
	\label{eq:loss_rate_1d}
\end{equation}

In PIC simulations,
the two-dimensional number density of the fast particle sheet
introduced above, $\sigma$,
is usually taken to be equal to
that of the plasma electron sheets;
in other words,
all particles in simulation have the same weight.
In this case, we have
\begin{equation}
	\sigma = \frac{n}{N_0^{(1)}},
\end{equation}
where $N_0^{(1)}$ is the mean one-dimensional (in $x$ direction) number density of the plasma electron sheets
and then Eq. (\ref{eq:loss_rate_1d}) is rewritten as
\begin{equation}
	\frac{dE}{dt} = - \frac{(q_0/e)^2}{2 N_e^{(1)}} \left|\frac{v_{0x}}{c}\right| \omgpe m_ec^2,
	\label{eq:loss_rate_1d_b}
\end{equation}
where $N_e^{(1)} = N_0^{(1)} \lmde$ is
the number of the plasma electron sheets contained in the electron skin depth $\lmde$.
Note that
for nonrelativistic fast particles
the equation (\ref{eq:loss_rate_1d_b}) can be rewritten in terms of the velocity of the fast particle as
\begin{equation}
	\frac{d|v_x|}{dt} = - \frac{1}{2 N_e^{(1)}} \frac{(q_0/e)^2}{m/m_e} \omgpe c,
	\label{eq:velocity_drag}
\end{equation}
which is exactly equivalent to the expression for the polarization velocity drag
for nonrelativistic fast electrons
obtained by \citet{Dawson62} (see also \cite{Eldridge63, Birdsall}) if we set $q_0=-e$ and $m=m_e$.

\subsection{Two dimensions (2D3V)}
As in the one-dimensional case,
the charge density of the fast particle in two dimensions (2D3V)
must be regarded as that made by a `rod' of the fast particles
in three dimensions.
Taking the simulation plane on the $x$-$y$ plane,
the identical fast particles are uniformly distributed in the $z$ direction
with the same $x$ and $y$ coordinates
and the charge density is given by
\begin{equation}
	\rho(\xv, t) = q_0 \eta \delta(x-v_{0x}t) \delta(y-v_{0y}t),
\end{equation}
where $\eta$ is the one-dimensional number density of the fast particles
embedded in the rod.
Thus,
the equation (\ref{eq:loss_rate}) becomes
\begin{equation}
	\frac{dE}{dt} = \frac{q_0^2 \eta v_{0xy}}{ \pi} \int_\kmin^\kmax dk \int_0^{2\pi} d\theta \cos\theta \mathrm{Im}\left[\epsilon(\kv, kv_{0xy}\cos\theta)\right]^{-1},
\end{equation}
where $v_{0xy} = (v_{0x}^2 + v_{0y}^2)^{1/2}$.
Using the approximation (\ref{eq:inv_epsilon_approx2}) and assuming $L > 2\pi \lmde$,
we have
\begin{equation}
	\frac{dE}{dt} = -2 q_0^2 \eta \omgpe \arccos(\omgpe/\kmax v_{0xy}).
\end{equation}
Then, assuming $\kmax \gg\omgpe/v_{0xy}$,
we finally obtain
\begin{equation}
	\frac{dE}{dt} = -\pi q_0^2 \eta \omgpe.
\end{equation}
This is independent of the velocity
even for the nonrelativistic particles unlike the one dimensional case (\ref{eq:loss_rate_1d}).

When the fast particle has the same weight as the plasma electrons,
we have
\begin{equation}
	\eta = \frac{n}{N^{(2)}_0},
\end{equation}
where $N^{(2)}_0$ is the mean two-dimensional number density of the plasma electron rods
and then we obtain
\begin{equation}
	\frac{dE}{dt} = - \frac{(q_0/e)^2}{4 N_e^{(2)}} \omgpe m_ec^2,
	\label{eq:loss_rate_2d_b}
\end{equation}
where $N_e^{(2)} = N_0^{(2)} \lmde^2$ is the number of the plasma electron rods
contained in the area of the electron skin depth squared, $\lmde^2$.
For nonrelativistic fast particles,
as Eq.~(\ref{eq:velocity_drag}) in the one-dimensional case,
this equation can be rewritten as the velocity drag form as
\begin{equation}
	\frac{dv_{0xy}}{dt} = - \frac{1}{4 N_e^{(2)}} \frac{(q_0/e)^2}{m/m_e} \frac{c}{v_{0xy}} \omgpe c.
\end{equation}
%

\subsection{Effects of multi-components, shape factors and spatial filters}
In PIC simulations,
particles are not point charges,
but have a shape with finite extension comparable to the grid size $\Delta x$.
The shape of the particles is expressed by the shape factor $S(\xv)$ \cite{Birdsall, Hockney}.
For example, in one dimension,
the shape factor of the cloud-in-cell (CIC) model is given by
\begin{equation}
	S(x) =
		\left\{ 
		\begin{array}{ll}
			1-|x|/\Delta x & (|x|/\Delta x < 1)\\
			0 & (|x|/\Delta x > 1).
		\end{array}
		\right.
\end{equation}
The charge density may also be smoothed out by using a spatial filter
with a kernel $S_F(\xv)$.
Furthermore,
PIC simulations generally deal with multi-component plasmas,
not single-component plasmas.
Therefore,
the dielectric response function is altered accordingly.
Within the grid less model,
the dielectric response function for a multi-component plasma,
where the mass and the temperature of the species $s$ are given by
$m_s$ and $T_s$ respectively,
with the effects of the shape factor and the spatial filter
may be given by \cite{Birdsall}
\begin{equation}
	\epsilon(\kv, \omega) = 1 + \sum_s \frac{\kDs^2}{k^2} |S(\kv)|^2 |S_F(\kv)|^2 W\left(\frac{\omega}{k\vths}\right),
	\label{eq:DRF_multi}
\end{equation}
where $\vths = (T_s/m_s)^{1/2}$, $\kDs = \omgps/\vths$ and $\omgps$ are
the thermal velocity, the Debye wavenumber and the plasma frequency of the $s$ species, respectively.
$S(\kv)$ is the Fourier transforms of the shape factor
and $S_F(\kv)$ is that of the spatial filter.
However, these effects are usually negligible as shown below.

For ions with $m_s \gg m_e$ and the same temperature $T_s = T_e$,
the argument of the W function for the s species,
$Z_s = v_0 \mu/\vths \propto \sqrt{m_s}$,
is much larger than that for the electrons.
Therefore,
noting Eqs. (\ref{eq:W_imag}) and (\ref{eq:W_approx}),
we see that
the effect of the ions in (\ref{eq:DRF_multi}) is usually negligible.
The only exception is the electron-positron plasma;
in this case the energy loss rate is obtained
by simply replacing $\omgpe^2$
with $2 \omgpe^2$ in the expressions for the electron plasma
because of the symmetry of the two species.
(However, note that the energy loss rates derived here is valid for the nonrelativistic plasmas
because we have assumed the nonrelativistic Maxwellian distribution
in Eqs. (\ref{eq:DRF}) and (\ref{eq:DRF_multi}).)

In addition,
as shown in Eq.~(\ref{eq:inv_epsilon_approx}) or Eq.~(\ref{eq:inv_epsilon_approx2}),
the function $\mathrm{Im}[\epsilon(\kv,\kv\cdot\vv_0)]^{-1}$
takes non-zero values only around $|\kv| \sim \omgpe/v_0$
for sufficiently fast particles.
Therefore,
the values of $S(\kv)$ and $S_F(\kv)$
only at that wavenumber are relevant to the energy loss rate.
However,
they are approximately unity at that wavenumber,
$S(\omgpe/v_0) \sim S_F(\omgpe/v_0) \sim 1$,
because both $S(\xv)$ and $S_F(\xv)$ have the spatial sizes of
the order of the grid size $\Delta x$
and usually the condition $\Delta x \ll v_0/\omgpe = (v_0/c) \lmde$ is satisfied.
Therefore, these effects are also negligible for ordinary cases.

Thus,
even for multi-component plasmas with
the shape factor and/or the spatial filter,
the energy loss rates can be approximated by
that derived for the electron plasma
very well for ordinary situations.

\subsection{Normalized expressions and dimensional dependence}
Rewriting the equations (\ref{eq:loss_rate_1d_b}),  (\ref{eq:loss_rate_2d_b})
and (\ref{eq:loss_rate_3d_d}) in terms of
the normalized quantities,
$E' = E/m_ec^2$, $t' = \omgpe t$, $q_0' = q_0/e$, $v_0'=v_0/c$ and $v_{0x}'=v_{0x}/c$,
we obtain
\begin{equation}
	\frac{dE'}{dt'} = 
	\left\{
		\begin{array}{rl}
			\displaystyle
			-\frac{q_0'^2}{2 N_e^{(1)}} |v_{0x}'|  &  \textrm{(for 1D)}\\
			\displaystyle
			- \frac{q_0'^2}{4 N_e^{(2)}}  &  \textrm{(for 2D)}\\
			\displaystyle
			-\frac{q_0'^2}{4\pi N_e^{(3)}} \frac{1}{v_0'} \ln(\kmax v_0/\omgpe)  &  \textrm{(for 3D)}.
		\end{array}
	\right.
	\label{eq:energy_loss_rate}
\end{equation}
We see that
the fundamental parameter for the energy loss rate of fast particles
is the number of electrons contained in the corresponding
electron skin depth volume; $N_e^{(1)}$, $N_e^{(2)}$, or $N_e^{(3)}$ for respective dimensionality.
For relativistic fast particles ($v_0 \sim c$),
the order of the energy loss rate in this unit is simply given by $\sim q_0'^2/ N_e^{(d)}$
for all cases.
For nonrelativistic fast particles ($v_0 < c$),
there are different dependences on the velocity for different dimensionality.

Summarizing the above argument,
we note that
for the same numbers of electrons in the skin depth volume, $N_e^{(d)}$,
the energy loss rate of the relativistic fast particles is effectively
independent of the mass and the energy of the fast particles,
the temperature of the plasma, and
the number of electrons in the volume defined by the Debye length.
It is also almost independent of the composition of the plasma,
the shape factors and the spatial filters used in the PIC simulation for ordinary cases
as shown in the previous subsection.

\section{Numerical Experiments}
To confirm the theoretical results derived in the previous section,
we show here the results of some numerical experiments
mainly with one dimensional (1D3V) simulations.

In the following, we consider the energy loss of a fast electron with $m=m_e$ and $q_0=-e$
moving through a uniform plasma in the $x$-direction.
Basic parameters and settings of the simulations are as follows:
the initial four-velocity of the fast electron $u_{0x}=10$
which corresponds to the three-velocity of $v_{0x}=0.995c$ and
the kinetic energy of $E=9.05 m_ec^2$, 
the ion-to-electron mass ratio of the plasma $m_i/m_e=20$,
the temperature of the plasma $T=16$keV ($\vthe=0.177c$),
the grid size $\Delta x = 0.1\lmde$,
the time step $\Delta t=0.05\omgpe^{-1}$,
and
the length of simulation system $L=409.6\lmde$.
The cloud-in-cell (or the first-order) shape factor is used
and no spatial filters are adopted.
Several of these parameters are altered in the following experiments
in order to investigate the dependences on them.
The 1D3V simulation code used is a second-order finite-difference code
with a staggered grid and semi-implicit scheme for the electromagnetic field
developed based on \cite{Birdsall}.
Initially,
the plasma is uniform and isothermal,
and the periodic boundary conditions are imposed at the ends of the simulation box.

\begin{figure}[htbp]
\begin{center}
	\includegraphics[width=11cm]{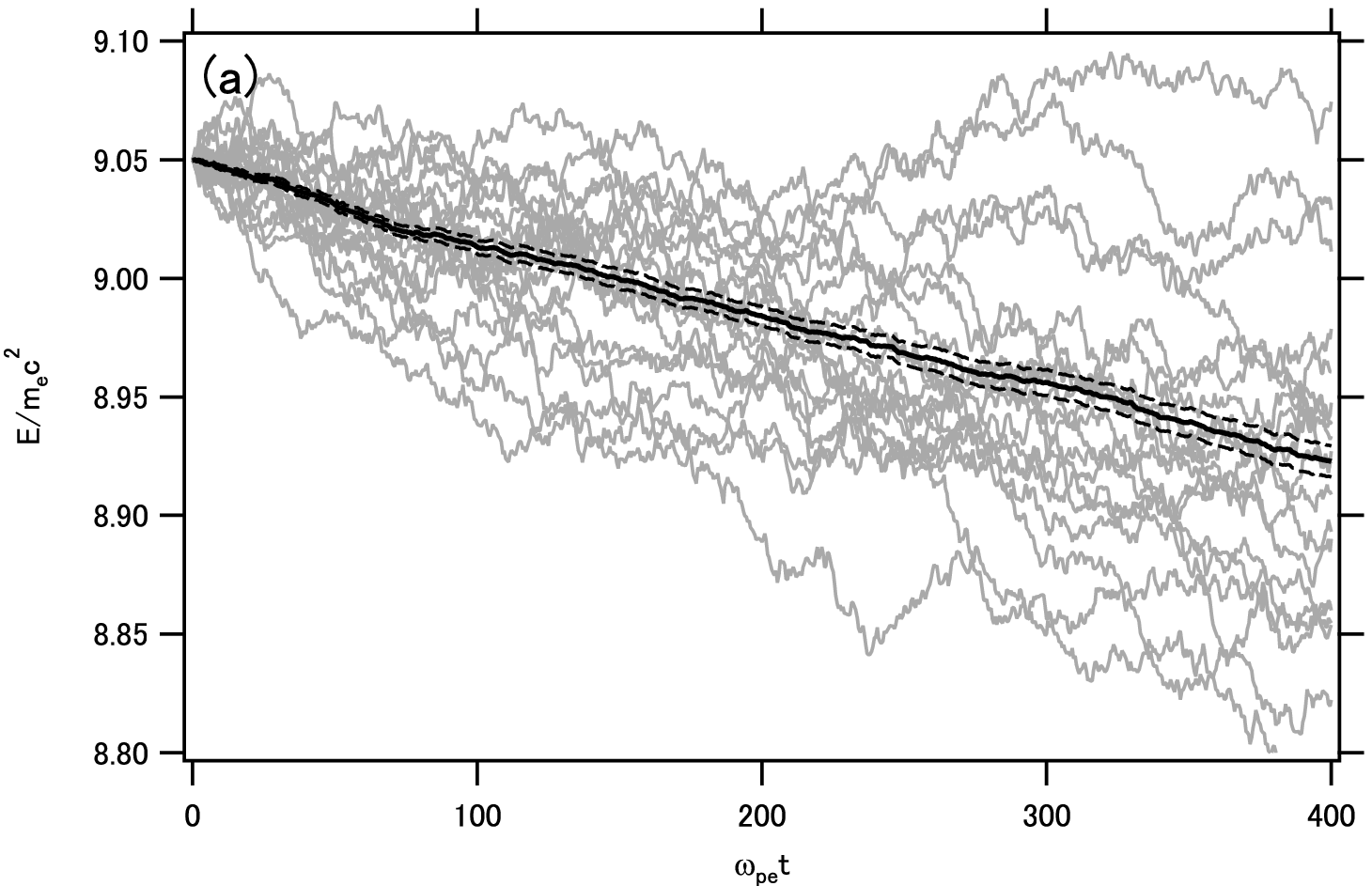}\\
	\includegraphics[width=11cm]{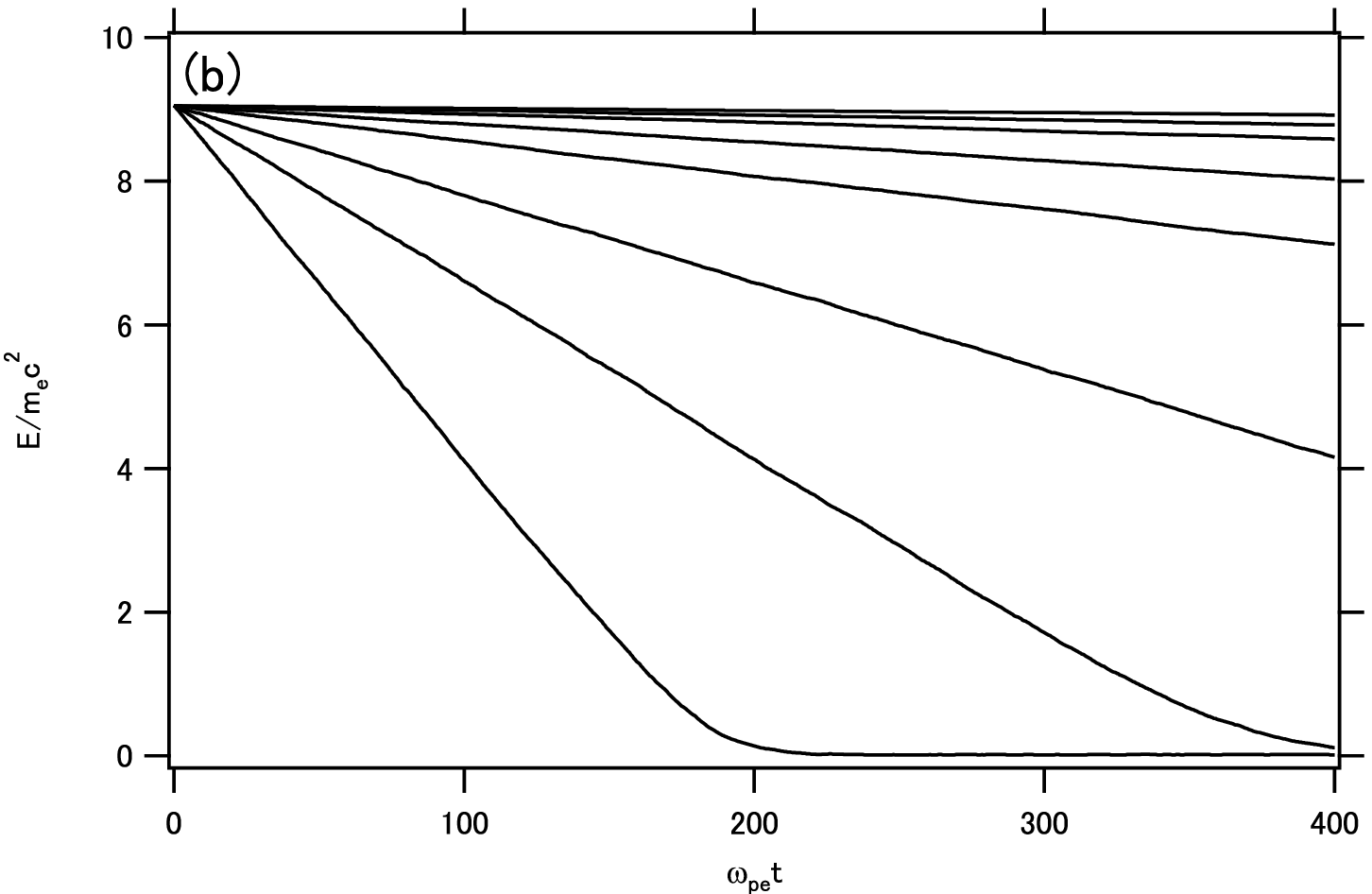}
\end{center}
\caption{
Energy history of a fast particle with $u_{0x}=10$ obtained from 1D3V simulations.
(a) Energy histories of the fast particle for a fixed number of plasma electrons in the electron skin depth $N_e^{(1)}=1600$. Those obtained from twenty independent runs
are shown with the light gray curves.
The average energy history calculated from one hundred runs is also shown by the black solid curve.
The upper and lower dashed curves denote the 1-$\sigma$ confidence interval of the average value.
(b) The average energy histories for various  $N_e^{(1)}$ of (from top to bottom) $1600, 800, 400, 200, 100, 40, 20$ and $10$ are shown.
}
\label{fig:fig1}
\end{figure}

First,
we observe the energy loss process of a relativistic fast electron
with the 1D3V simulations.
Since the energy loss process is a stochastic one,
a number of energy histories of the fast particle are calculated by
independent simulations with the identical parameters and different random seeds
and then the average history is calculated from them.
The results from the total of one hundred runs are summarized in Fig. \ref{fig:fig1} (a).
The twenty of the individual energy histories are shown by the light gray curves
and the average of the all histories is shown by the black solid curve.
Although the individual histories are generally highly fluctuating,
their average obeys almost a linear evolution (i.e., $dE'/dt'=\mathrm{const}$)
as expected from Eq. (\ref{eq:energy_loss_rate}).
Figure \ref{fig:fig1} (b) shows
the average energy histories calculated
for various numbers of the plasma electrons contained in the electron skin depth, $N_e^{(1)}$.
The dependence of the average energy loss rates on $N_e^{(1)}$ is evident.

\begin{figure}[htbp]
\begin{center}
	\includegraphics[width=11cm]{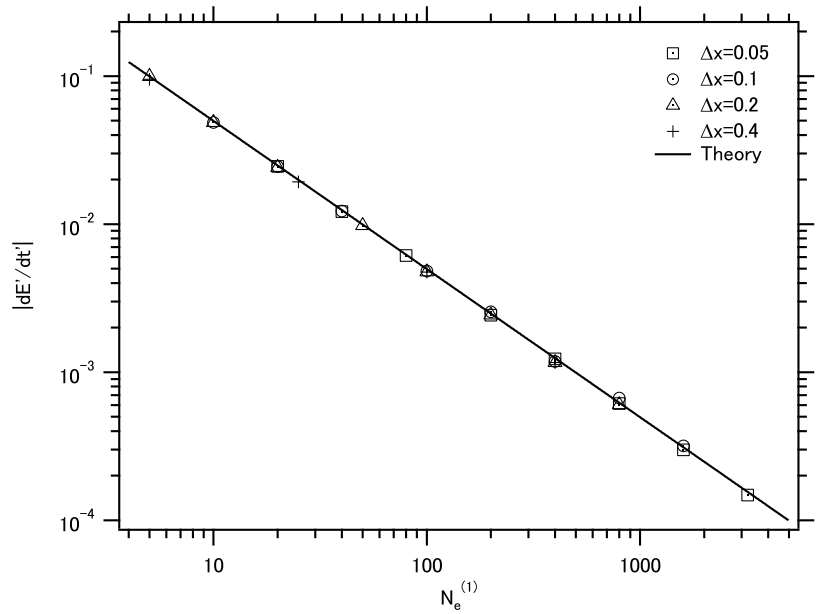}
\end{center}

\caption{
Dependence of the average energy loss rate on the number of plasma electrons
contained in the electron skin depth length, $N_e^{(1)}$,
and on the grid size $\Delta x$.
The simulation parameters other than $N_e^{(1)}$ are the same as those in Fig.\ref{fig:fig1}.
The simulation results and the theoretical result (\ref{eq:energy_loss_rate})
are shown by the symbols and by the solid curve, respectively.
}
\label{fig:1d_ux10_T16000}
\end{figure}

Figure \ref{fig:1d_ux10_T16000} shows the dependence of the average energy loss rate on $N_e^{(1)}$
and on the grid size $\Delta x$.
The average energy loss rates are calculated from the simulation results
by averaging a number of individual energy histories as in Fig. \ref{fig:fig1} (a)
and then fitting the straight portion of the average histories.
We also calculated the 1-$\sigma$ confidence interval of each average value,
which depends on the number of simulations carried out to obtain the average.
However,
we omit them in the figure
because the error bars are smaller than the size of the symbols.
We see that the simulation results 
almost perfectly fit the theoretical curve given by Eq.(\ref{eq:energy_loss_rate}).
This figure also demonstrates that
the energy loss rate does not depend on the grid size $\Delta x$,
or equivalently on the number of particles per cell $N_{PPC}$,
when $N_e^{(1)}$ is fixed.

\begin{figure}[htbp]
\begin{center}
	\includegraphics[width=11cm]{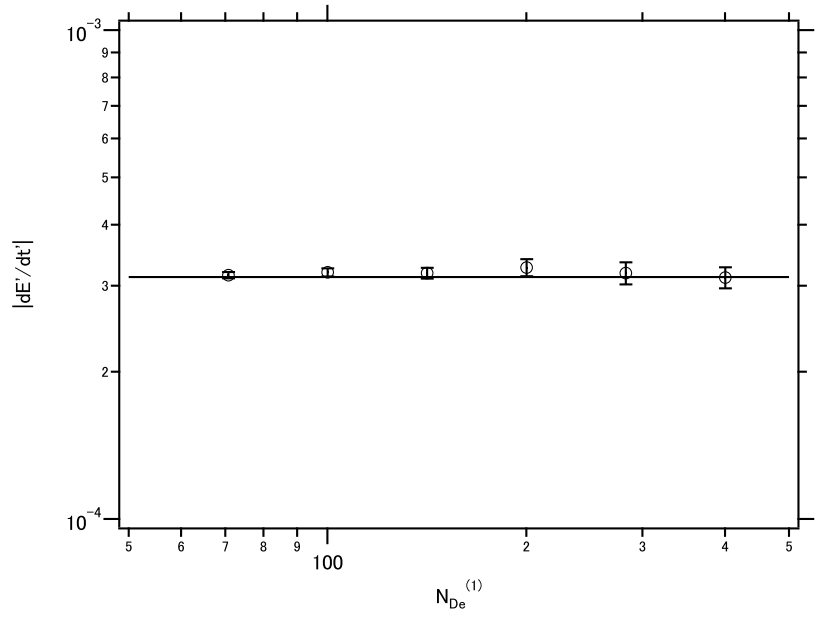}
\end{center}

\caption{
Dependence of the energy loss rate on the number of plasma electrons
contained in the Debye length, $N_\mathrm{De}^{(1)}$,
for fixed $N_e^{(1)} = 1600$.
Symbols denote the simulation results with error bars for the 1-$\sigma$ confidence intervals.
Corresponding temperatures of the plasmas are, from left to right, 1, 2, 4, 8, 16 and 32 in keV,
respectively.
The theoretical prediction is shown by the solid curve.
}
\label{fig:1d_ux10_Temp}
\end{figure}

Figure \ref{fig:1d_ux10_Temp} shows the dependence of the energy loss rate
on the number of plasma electrons contained in the Debye length, $N_\mathrm{De}^{(1)}$,
or equivalently on the temperature of the plasma, for fixed $N_e^{(1)} = 1600$.
The other parameters are the same as in Fig. \ref{fig:1d_ux10_T16000}.
We see that the energy loss rate does not depend on $N_\mathrm{De}^{(1)}$
when $N_e^{(1)}$ is fixed.
This fact is interesting because
the collision frequency of thermal energy particles in PIC simulations
is basically determined by $N_\mathrm{De}^{(1)}$ \cite{Hockney, Hockney71}, not by $N_e^{(1)}$.

\begin{figure}[htbp]
\begin{center}
	\includegraphics[width=11cm]{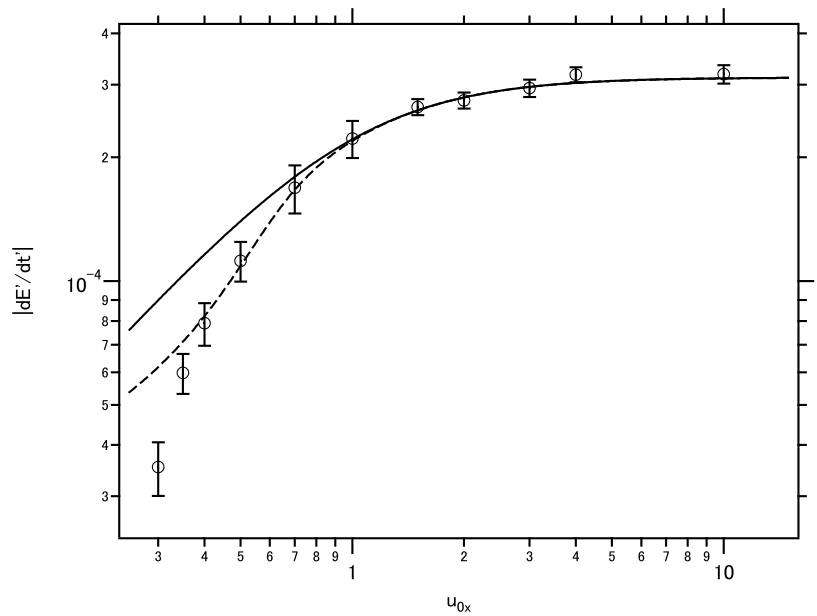}
\end{center}

\caption{
Dependence of the energy loss rate on the velocity of the fast particle for fixed $N_e^{(1)} = 1600$.
The symbols denote the simulation results and the solid curve shows the theoretical curve
given in (\ref{eq:energy_loss_rate}).
The dashed curve shows another theoretical curve using a better approximation (see text).
}
\label{fig:1d_ux_dependence}
\end{figure}

\begin{figure}[htbp]
\begin{center}
	\includegraphics[width=11cm]{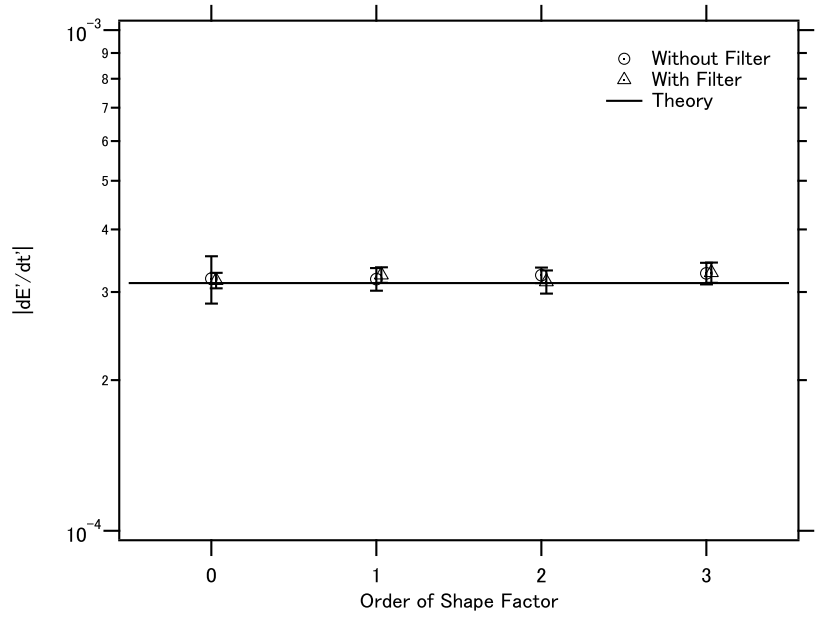}
\end{center}

\caption{
Dependence of the energy loss rate on the shape factors and the spatial filters for fixed $N_e^{(1)} = 1600$.
The energy loss rates obtained from the simulations are shown by the symbols for the shape factors of,
from left to right, zeroth-, first-, second- and third-order.
The circles and the triangles respectively denote those without and with the spatial filters.
}
\label{fig:1d_ux10_shape_filter}
\end{figure}

Figure \ref{fig:1d_ux_dependence} shows the dependence of the energy loss rate
on the velocity of the fast particle.
The solid curve shows the theoretical result given in (\ref{eq:energy_loss_rate}).
The dashed curve shows another theoretical curve
using a better approximation for small $u_{0x}/\vthe$
obtained from Eqs. (\ref{eq:loss_rate_1d_0}) and (\ref{eq:W_approx}).
The dependence on the velocity is evident for the nonreletivistic regime.
The simulation results are in agreement with the theoretical curve of the latter one well
except for small values of $u_{0x}$;
For $T=16$keV ($\vthe=0.177c$) and $u_{0x}=0.3$ ($v_{0x}=0.287c$), the argument of the W function becomes $Z \sim 1.62$ and
the large $Z$ approximation gets worse for that case.
Figure \ref{fig:1d_ux10_shape_filter} shows the dependence on the shape factors and the spatial filters.
Here,
a simple three-point filter is used (see Appendix C in \cite{Birdsall} for details).
It is evident that the energy loss rate is independent of them as mentioned in Sec.2.4.

\begin{figure}[htbp]
\begin{center}
	\includegraphics[width=11cm]{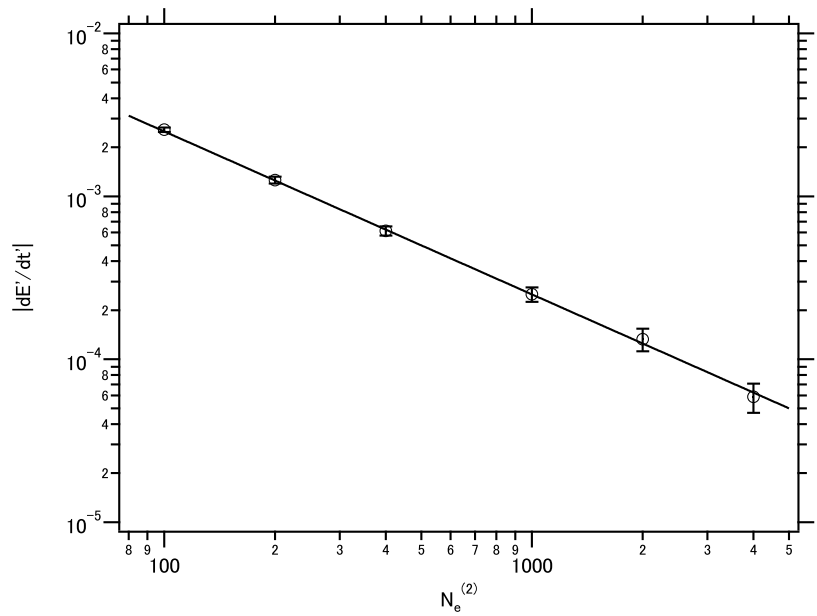}
\end{center}

\caption{
The average energy loss rate in two-dimensional (2D3V) simulation as a function of $N_e^{(2)}$.
The symbols denote the results from simulations while the solid curve shows
the theoretical result given in (\ref{eq:energy_loss_rate}).
The error bars show the 1-$\sigma$ confidence intervals of the average values.
}
\label{fig:2d}
\end{figure}

Figure \ref{fig:2d} shows the average energy loss rate in two-dimensional (2D3V) simulations
as a function of $N_e^{(2)}$.
The simulation code uses the spectral method and the third-order shape factor.
(However,
the energy loss rate should be independent of the details of the simulations
if the argument presented in the previous section is true.)
We confirm that
the simulation results are in agreement with the theoretical curve (\ref{eq:energy_loss_rate})
very well.

\section{Concluding Remarks}
In this paper,
we have shown fast particles in PIC simulations
are suffered from the energy loss
due to the stopping power of a plasma
and the energy loss rate of relativistic fast particles
is given by $\sim q_0'^2/N_e^{(d)}$ in the normalized unit.
However,
since the number $N_e^{(d)}$ in PIC simulations
is usually much smaller than that in real plasmas,
the energy loss can be exaggerated significantly
and can affect the simulation results.
Therefore,
especially for the simulations that investigate the particle acceleration processes
that work in relatively long timescales,
the number $N_e^{(d)}$
should be chosen large enough to avoid this artificial energy loss.

For example,
for the first-order Fermi acceleration in collisionless shocks \cite{Blandford87},
the mean energy gain rate is roughly given by
\begin{equation}
	\left( \frac{dE'}{dt'} \right)_\mathrm{acc}
	= \frac{3}{20} \frac{q_0'}{\xi} \left( \frac{V_s}{c} \right)^2 \frac{\omgce}{\omgpe}
	\label{eq:Fermi_energy_gain}
\end{equation}
in the normalized unit,
where $\omgce$ is the electron cyclotron frequency, $V_s$ is the shock velocity
and $\xi$ is the parameter in the Bohm diffusion model in which
the mean free path of the particles being accelerated is given by
$\sim \xi r_g$ where $r_g$ is the gyroradius of particle \cite{Jokipii87}.
Although the value of $\xi$ is unknown a priori
and should be determined by the condition
of the magnetic irregularity in the system,
the values of $\xi \sim 1-10$ are often assumed for efficient acceleration.
In Eq.(\ref{eq:Fermi_energy_gain}),
the shock compression ratio of 4 is also assumed.
To investigate the acceleration process with PIC simulations,
the condition
\begin{equation}
	\left(\frac{dE'}{dt'}\right)_\mathrm{acc}
	 > \left( \frac{dE'}{dt'} \right)_\mathrm{loss}
\end{equation}
must be satisfied,
where $(dE'/dt')_\mathrm{loss}$ is the energy loss rate given in (\ref{eq:energy_loss_rate}).
Thus,
for instance, for one-dimensional simulations,
the number of the plasma electrons contained in the electron skin depth length, $N_e^{(1)}$,
must satisfy
\begin{equation}
	N_e^{(1)} \gg \frac{10}{3} q_0' \xi \left(\frac{V_s}{c}\right)^{-2} \left(\frac{\omgce}{\omgpe}\right)^{-1}.
\end{equation}
Otherwise,
the exaggerated energy loss may make the acceleration process inefficient.




\newpage

\newpage

\newpage

\newpage

\newpage

\newpage

\end{document}